# Real-time observation of polyelectrolyte-induced binding of charged bilayers

*Yuxia Luan,* [1] *and Laurence Ramos\**

LCVN (UMR CNRS-UM2 n°5587), CC26, Université Montpellier II, 34095, Montpellier Cedex 5, France

E-mail: ramos@lcvn.univ-montp2.fr

*Abstract:* We present real-time observations by confocal microscopy of the dynamic behavior of multilamellar vesicles (MLVs), composed of charged synthetic lipids, when put in contact with oppositely charged polyelectrolyte (PE) molecules. We find that the MLVs exhibit astonishing morphological transitions, which result from the discrete and progressive binding of the charged bilayers induced by a high PE concentration gradient. Our physical picture is confirmed by quantitative measurements of the fluorescence intensity as the bilayers bind to each other. The shape transitions lead eventually to the spontaneous formation of hollow capsules, whose thick walls are composed of lipid multilayers condensed with PE molecules. This class of objects may have some (bio)technological applications.

---

[1] Permanent address: Institute of Pharmaceutics, School of Pharmacy, Shandong University, Jinan 250012, China

# 1 Introduction

Liposomes are often studied as simplified models of biological membranes [1,2] and are extensively used in the industrial area ranging from pharmacology to bioengineering.[3] The biomimetic properties of the membrane also make liposomes attractive as vessels for model systems in cellular biology.[4,5] Composite systems of lipid bilayer and polymers have received special attention due to their similarity with living systems such as plasma membrane and various organelle membrane, that mainly consist of complex polymers and lipids.[6] Experimental investigations of vesicle/polymer mixed systems also aim at improving the stability and at controlling the permeability of liposomes for drug delivery or targeting, or for gene therapy.[7-12] For example, stabilization is usually obtained by loose hydrophobic anchoring of water soluble chains that do not significantly perturb the bilayer organization, such as alkyl-modified Dextran or Pullulan with a low degree of substitution,[13,14] long poly(ethylene glycol) capped with one or two lipid anchors per macromolecule, or poloxamers.[15,16] It was shown recently[7,17-20] that water-soluble polymers, upon binding to vesicles, can markedly affect the shape, curvature, stiffness, or stability of the bilayer. However, the mechanisms of these polymer-induced reorganizations of membranes remain sometimes conjectural, although it is clear that the hydrophobicity of the polymer plays an important role.

On the other hand, interactions between surfactants and polymers in bulk solution are extensively investigated, due to their numerous applications from the daily life to the various industries (e.g. pharmaceutical, biomedical application, detergency, enhanced oil recovery, paints, food and mineral processing).[21-25] Charged amphiphilic molecules, like lipids or surfactants, and oppositely charged polyelectrolytes (PE) spontaneously form stable complexes, which are very promising objects, because of their great variability in structures and properties.[26-27] In this context, interactions of charged bilayers with oppositely charged PE are particularly regarding. For instance, in bioengineering, the interactions between lipids bilayers and DNA molecules are crucial for gene therapy.[28-32] When charged bilayers interact with polyelectrolyte of opposite charge, it is generally accepted that electrostatic interactions induce the bridging of the lipid bilayers by the PE molecules.[33-35] The resulting structure of the PE/lipid complexes is a condensed lamellar phase, with PE strands intercalated between the lipids bilayers. However, although most studies provide a general picture for the PE/lipid structure, very few addressed the question of the mechanism of the formation of the complexes or the associated issue of dynamics and intermediate steps for the assemblage process. In addition to our previous work [36], two noticeable exceptions include the work of Kennedy *et al.* who found that the order of addition of DNA to cationic lipid or vice versa could affect the size and size distribution of the complexes [31] and that of Boffi *et al.* who showed that two distinct types of DNA/lipid complexes can be formed depending on the sample preparation procedure.[32] Nevertheless, the determinants for the assembly and dynamics of complex formation remain poorly understood.

Under certain conditions, lipids can self-assemble into giant vesicles, the size of living cells. These are very elegant objects that allow manipulation and real time observation with a light microscope [37-40] and that have opened the way to a wealth of theoretical and experimental investigations.[41] However, unlike experimental work on giant unilamellar vesicles, experimental reports on real-time observation of the effect of a chemical species on the stability and shape changes of a multilamellar vesicles are extremely scarce.[42-46] Nevertheless, as it is demonstrated in this paper, when multilamellar vesicles are used, richer behaviours can be expected, since cooperative effects due to the dense packing of bilayers may play an important role.

In the present study, we employ a real-time approach to study the dynamics of the interactions between charged membranes and oppositely charged PE molecules, and monitored by light and confocal microscopy the behavior of multilamellar vesicles (MLVs) made of a synthetic lipid in a concentration gradient of PE. When the gradient is strong enough, the MLV undergoes spectacular morphological transitions, which enable us to visualize the progressive binding of charged bilayers induced by oppositely charged PE molecules. Specifically, these shape transitions lead eventually to the spontaneous formation of a hollow capsule with thick walls that are presumably composed of lipid multilayers condensed with PE molecules. This class of objects may have some potential (bio)technological applications[47] and this contribution could have some significance in mimicking bioprocess. We first present our experimental observations, then describe the mechanisms at play and provide quantitative measurements, based on the fluorescence intensity, which support our physical picture. Finally, we briefly conclude.

**2 Experimental Results**

We use vesicles made of didodecylammonium bromide (DDAB) as synthetic lipid, and an alternating copolymer of styrene and maleic acid in its sodium salt form, as anionic polyelectrolyte (PE). The DDAB bilayers are labeled with a fluorescent surfactant for confocal and fluorescence imaging. We follow by light and confocal microscopy the behavior of the DDAB vesicles when they are submitted to a PE concentration gradient. The Materials and Methods are described in the Supporting Information.

**2.1 Interaction between Giant Unilamellar Vesicles and Polyelectrolyte**

The time-dependent morphological changes of a giant unilamellar vesicle (GUV) are investigated when the GUV is exposed to a concentrated PE solution (30% W/W). The GUV is floppy and fluctuating before interacting with PE. Upon contact with the polyelectrolyte solution, the bilayer becomes tense and the vesicle immediately turns to perfectly spherical and taut. Some patches, that appear very intense in fluorescence, gradually formed on the surface of the GUV. The patches thicken with time (Figure 1). Concomitantly, the size of the GUV decreases. These processes lead ultimately to the collapse of the GUV, resulting in a single small lump made of a compact DDAB/PE complex. The duration of the whole process is of the order of several minutes. Analogous observations have been recently reported for the interaction of GUVs with small unilamellar vesicles[48], with the matrix protein of a virus[49], and with a flavonoid of green tea extracts[50].

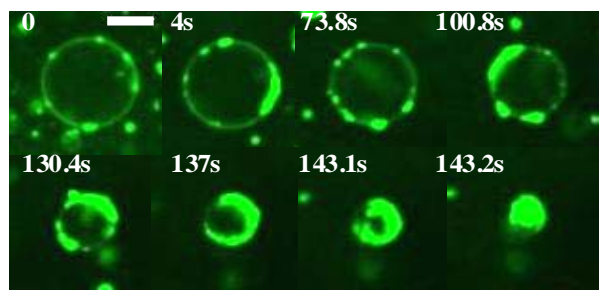

*Figure 1*. Evolution of the morphology of a giant unilamellar vesicle (GUV) upon contact with a concentrated PE solution (30% W/W). Timing is indicated in white text. The scale is the same for all pictures. Scale bar = 5 μm.

## 2.2 Interaction between Multilamellar Vesicles and Polyelectrolyte

### 2.2.1 Phase-diagram

In sharp contrast to the case of GUVs, the interactions of charged multilamellar vesicles (MLVs) with polyelectrolyte molecules of opposite charge lead to unexpectedly rich phenomena. We interestingly notice that, depending on $C_{PE}$, the PE concentration, completely different morphological transitions are observed. The "phase" diagram shown in Figure 2 summarizes our experimental findings for MLVs put into contact with different concentrations of PE. Successive peeling events are found when $C_{PE} < \sim 2\%$ as shown in Figure 2A, while concentrated PE ($C_{PE} > \sim 10\%$) induces the appearance of spectacular morphological changes of the MLV (Figure 2B and 2C). We confirmed by differential interference contrast and phase contrast microscopy that non-fluorescent MLVs exhibit identical morphological changes, and that all dynamical processes reported below are preserved.

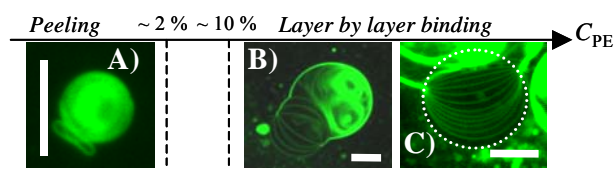

*Figure 2*. "Phase" diagram of MLV in contact with different PE concentrations, viewed by confocal imaging. Scale bars =10 μm.

### 2.2.2 Weak Polyelectrolyte Gradient

When a MLV is exposed to a diluted PE concentration, the size of the MLV gradually decreases and concomitantly small aggregates formed in the vicinity of the MLV. The MLV is peeled progressively, layer after layer, one DDAB/PE complex being formed for each peeling event, while the interior of the MLV remains always intact. Peeling events proceed until the MLV is completely used up. The final state of the MLV is a pile of small aggregates of size ranging from 2 to 10 μm. The whole consumption of a MLV through the peeling mechanism is a slow process that lasts more than 10 minutes, each peeling events lasting about tens of seconds (Figure 3). We note that the effect of a weak polyelectrolyte gradient on a MLV has been reported previously.[45] However, the novel confocal microscopy pictures given in Figure 3 show unambiguously a single event, which provides a compelling evidence for a peeling mechanism.

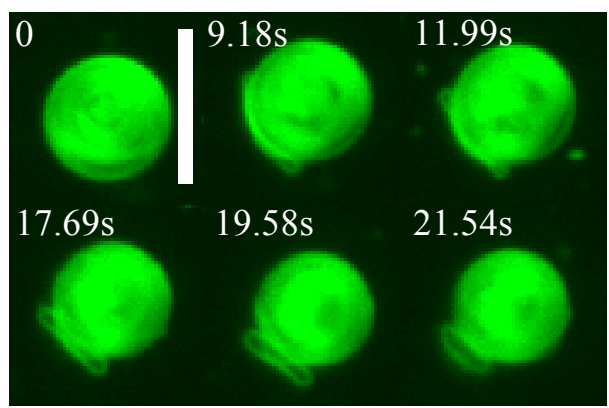

*Figure 3.* One peeling event of MLV induced by a diluted PE (0.5 % W/W). Timing is indicated in white text. The scale is the same for all pictures. Scale bar = 10 μm.

### *2.2.3 Strong Polyelectrolyte Gradient*

In sharp contrast with our observations for a dilute polyelectrolyte solution, for $C_{PE} > \sim 10\%$, the morphological transitions of a MLV lead to a finite-size cellular object, with water encapsulated in the cells, and whose walls are very likely made of DDAB/PE complexes (Figure 4E). The angles between the thick walls measured in 2-dimentional picture are about 120°, similarly to the angle at which film meet in a three-dimensional dry foam.[51] When the size of the initial MLV is sufficiently small, hollow capsules are eventually obtained (Figures 4B-D), whose size is sensibly equal to that of the initial MLV. Although the large scale structure depends dramatically on the initial PE concentration, the microscopic structure in all cases is a condensed lamellar phase (Figure 4G), as checked by small-angle X-ray scattering (Figure 4F), whose periodicity is of the order of 3.0 nm, hence only slightly larger than the bilayer thickness (2.4 nm).[52]

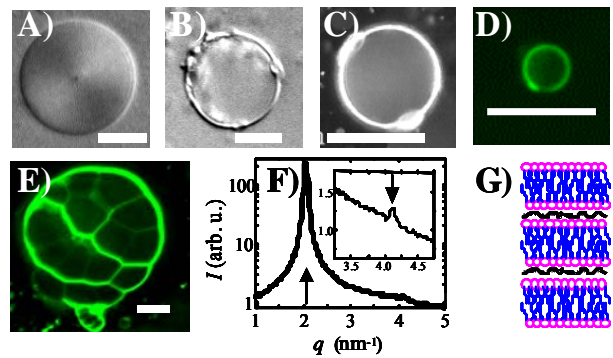

*Figure 4.* (A, B) Differential Interference Contrast; (C) Fluorescence and (D, E) Confocal imaging of, (A) a MLV prior to interaction with PE, (B, C, D) hollow capsules and (E) a cellular object, after interaction with a strong PE gradient. In (E), by comparison of the intensity of a single bilayer to that of the thickest wall, it is evaluated that the thickest external wall contains ~ 20 bilayers. Scale bars = 10 μm; (F) X-ray spectrum of the DDAB/PE complexes. The arrows point the peaks at $q_0$= 2.05 nm$^{-1}$ and $2q_0$, which indicates a lamellar phase; (G) Scheme of the microscopic structure of the complexes, consisting of a condensed lamellar phase with PE molecules intercalated between the bilayers.

The typical whole sequence of morphological transformation of a MLV when it is exposed to a concentrated PE solution is shown in the time series pictures of Figure 5. Before the MLV starts to deform significantly, the fluorescent intensity inside the vesicle becomes heterogeneous, the higher intensity being localized in the region with higher PE concentration. The surprising buds (Figure 2 B-C) composed of well-separated set of bilayers form subsequently. Interestingly, we note that the first striated buds form systematically where the PE concentration is lower. The interaction dynamics then speeds up and the MLV is found to experience rapid fluctuations, with the formations of protruding and budding, while dynamical events can also be distinguished in the core of the vesicle. The initially "full" MLV appears finally essentially devoid in DDAB bilayers: the inside of the resulting object is essentially black with some thick fluorescent strands. This cellular soft object forms therefore a peculiar kind of biliquid foam[53-55]. As opposed to our observations for a weak polyelectrolyte gradient, the dynamics is here very fast: the whole sequence lasts less than 1 minute.

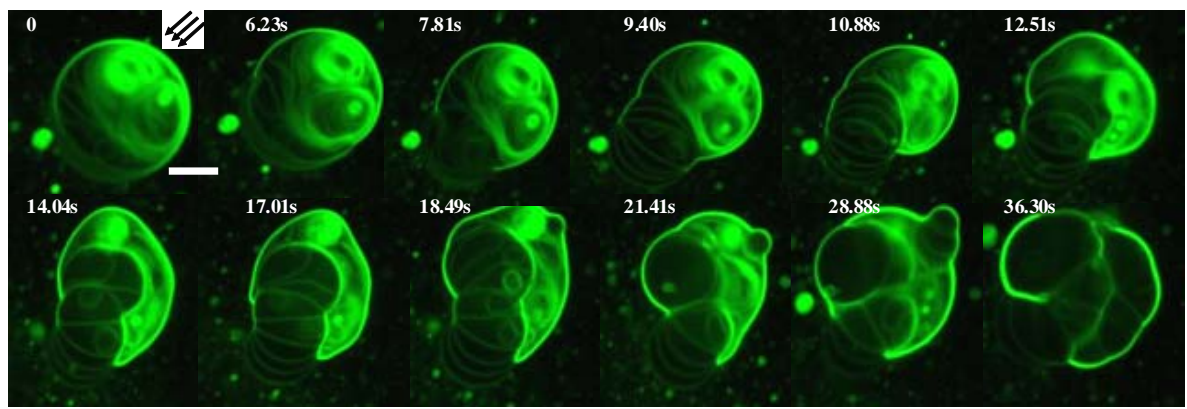

*Figure 5.* Time series showing the shape transformation of a MLV upon contact with a concentrated PE solution (30%W/W). The direction along which the polyelectrolyte molecules diffuse is shown with arrows. The whole process is shown, from the budding of the MLV, due to the layer by layer binding, to the formation of a cellular object. Timing is indicated in white text. The scale is the same for all pictures. Scale bar = 10 μm.

We finally note that we have performed some additional tests. First, we have done experiments in salted water (with NaBr) instead of pure water. The main observations described above are preserved with a salt concentration of $10^{-3}$ M. With a NaBr concentration of $10^{-2}$ M, our experimental observation in pure water cannot be reproduced due to the lack of stability of the MLVs.[56] Secondly, we have also investigated the interaction of MLV with other polymers, both neutral and charged (as listed in the Supporting Information), and found similar results as those described here only with the polystyrene-sulfonate polyanions, thus confirming that attractive electrostatic interactions between the DDAB bilayers and the polymer molecules are a key ingredient for our observations.

# 3 Discussion

Due primarily to the strong electrostatic interactions between charged bilayers and oppositely charged polyelectrolyte molecules, DDAB/PE complexes form, whose structure is a condensed lamellar phase.[36,45] The confocal pictures of a GUV interacting with a PE solution (Figure 1) provide a dynamic observation for the formation of these complexes. In this part, we discuss the experimental findings on the formation of DDAB/PE complexes, when polyelectrolyte molecules interact with a MLV. As we showed in the experimental section, depending on the PE concentration, the polyelectrolyte molecules interact with a unique bilayer (when the PE gradient is weak), or with the entire stack of bilayers (when the PE gradient is strong).

## 3.1 Interaction between PE and a unique bilayer

Upon contact with a weak PE gradient, a MLV is peeled off gradually. Each peeling event implies firstly the formation of a pore, which expands until failure of the entire bilayer. We have previously visualized the expansion of a pore by light microscopy.[45] Pore formation in unilamellar vesicles has been observed under different experimental conditions, including application of an electric field [57-58] interaction with proteins [37,50] or with a water-soluble polymer with hydrophobic pendent groups,[59] or attractive interactions with a patterned surface.[60] In our case, pores form because of the adsorption of PE onto the DDAB bilayers due to a strong electrostatic attraction between the two species. In fact, because of these interactions, part of the surface area of the external bilayer may be used up to form PE/lipid complexes. This creates a tension in the bilayer which ruptures above a critical tension, leading to the formation of a pore. The peeling mechanism was previously discussed in details.[45]

## 3.2 Interaction between PE and a stack of bilayers

### 3.2.1 PE-induced binding of two bilayers as elementary mechanism

We argue that the astonishing structures exhibited upon contact of a multilamellar vesicle with a strong gradient of PE concentration are due to a discrete and progressive binding of the bilayers induced by the PE molecules as they diffuse within the multilayer material. The elementary initial event can be imaged in real-time and is shown in Figure 6. It consists of the budding starting from the outmost bilayer. This structure results from the expulsion of the water that is located between the outmost and the secondary outer bilayers as they rapidly bind to each other due to the bridging of the oppositely charged PE between them. The binding front can be followed by confocal imaging: with time, the binding quickly spreads and the water between the bilayers is driven into a small and spherical water pool. Such events typically last a few seconds, and are faster when the PE concentration gradient is higher. A scheme of the microscopic process is shown in Figure 6E. We note that a temperature-induced binding of bilayers has been observed by light microscopy, but the dynamics could not be followed.[61] The succession of such events, i.e. binding of the secondary outer with the ternary outer bilayers, then binding of the ternary outer with the quaternary outer bilayers, … leads to the striated structures shown Figures 2B,C and 7. These structures originate from the successive formation of water pools, while the core of the MLV remains intact. The further interaction with PE leads to the binding of bilayers in the core of the MLV: the initially homogeneous contrast inside the MLV (Figures 2A, 4A) becomes progressively extremely heterogeneous as bilayers bind to each other and leave large portions free of bilayers. This is simultaneously accompanied by more important and erratic shape transformation, which leads ultimately to the formation of a cellular biliquid foam or hollow capsule (Figures 4 and 5).

More quantitatively, the volumes measured by image analysis can be compared with the volumes evaluated from the simple model (scheme, Figure 6E). We take for the water thickness between DDAB bilayers, 80 nm, the maximum swelling of the lamellar phase (prior to interaction with PE) [52] and calculate, for the MLV of Figure 6 (radius 10.7 µm), the volume of the water pool after binding of the outmost and secondary outer bilayer, $V_c$. We find $V_c$=130 µm$^3$. We compare $V_c$ to the volume $V_m$, for the water pockets evaluated from Figures 6C and D. We find $V_{m,6C}$ = 400 µm$^3$ ≈ $3V_c$ and $V_{m,6D}$ = 530 µm$^3$ ≈ $4V_c$, respectively, as expected since 3 and 4 elementary events have occurred respectively in C and D (as clearly distinguished in a movie of the process, movie S1 in SI). Similarly, we measure that the total water volume for 10 bilayers (white circle, Figure 2C) is about 4300 µm$^3$, while we calculate that the volume resulting from the binding of two bilayers is about 470 μm$^3$, hence roughly 10 times smaller, as expected. The very good agreement between the numerical values confirms the mechanism we propose, and suggests that there is no water release during this process.

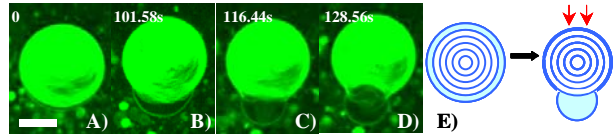

*Figure 6*. (A-D) Series of the morphological transformation of a MLV as it interacts with PE. The PE diffuses from top to bottom and $C_{PE}$ = 30%. The pictures show a succession of elementary events as schematized in (E). Timing is indicated in white text. Scale bar =10 μm

*3.2.2 Quantification of the discrete binding of the bilayers*

We follow the binding of individual surfactant bilayers into the thick bundles with confocal microscopy (Figure 7), and analyze the fluorescence intensity distribution with Image J. In Figure 8A, we show that the intensity profile, perpendicular to a bilayer, is homogeneous along the bilayer. We define *I*, the integrated intensity, as the surface area of the peak of the intensity profile. We found that *I* is constant for all individual bilayers (labeled *a* to *h*). The empty symbols in Figure 8C show *I* along the thick bundle P1-P2 (marked by the crosses). We measure that the intensity increases along the thick bundle from P1 to P2, which precisely reveals a discrete and continuous increase of *I* as more and more bilayers bind. To quantify this, we add to the intensity (along P1-P2) between bilayers *n* and *n*+1 the intensities of all bilayers with labels ranging from *n*+1 and *h* (*h* is the last bilayer). The calculated values are reported as full symbols in Figure 8C. Interestingly, we find that, at any step, these calculated intensities are a very good evaluation of the intensity (full hexagons) for the thickest part of the bundle (closed to P2). In fact, all full symbols Figures 8C are located on a same horizontal line. This provides a further and conclusive evidence of a discrete and progressive binding of bilayers.

Furthermore, our calculations demonstrate that the number of individual bilayers that compose a bundle can be evaluated from the fluorescent intensity of the bundle. For instance, by comparison of the intensity of a single bilayer to that of the thickest wall, we evaluate that for the cellular composite material (Figure 4E) the thickest external wall contains ~ 20 bilayers.

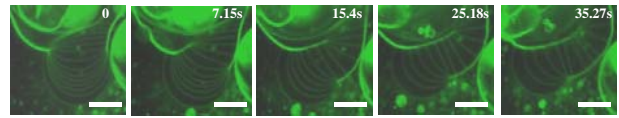

*Figure 7*. Pictures showing individual bilayers binding into a thick bundle. Scale bar =10 μm. Timing is indicated in white text.

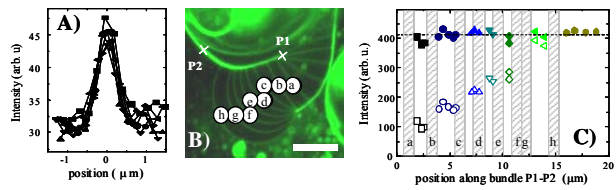

*Figure 8*. (A) Intensity profiles, performed at several positions along the bilayer *f*. (B) Binding of individual bilayers into a bundle (time series shown in Figure 7). Scale bar =10 μm. (C) Empty symbols: integrated intensity, *I*, along bundle P1-P2; full symbols: the sum of *I* the intensity along the bundle measured between the bilayers *n* and *n*+1 plus the intensities of all bilayers with labels ranging from *n*+1 and *h*, is very close to the final intensity (full hexagons), (closed to P2). The shaded area marks the locus where *I* cannot be measured because of the junction with an additional bilayer.

### 3.2.3 Kinetics is a key parameter for our observations

Importantly, we have noted that the events are polarized, the "budding" always occurring in the point diametrically opposed to the point where PE concentration is higher. This indicates that the binding always starts where PE concentration is higher and that the process is sensitive to the PE concentration gradient. In addition, the formation of buds indicates that the binding kinetics is faster than the diffusion of water across the compact bilayers. These experimental observations are consistent with the fact that the key parameter for this novel observation is to expose MLV to a strong gradient of PE. In addition, the hollow capsules formed have more or less the same size as the initial MLV when the initial MLV is not too large. This supports the fact that the water release, if any, is weak during the whole process, which is in full agreement with a binding kinetics faster than the diffusion of water across the bilayers.

In addition, our observations imply that the PE molecules penetrate inside the MLV. Very generally, the entry of the PE molecules into a MLV is driven by three forces: electrostatic interactions (between the surfactant headgroups and the maleic acid units of the PE), hydrophobic interactions (between the surfactant tails and the styrene units of the PE) and osmotic pressure (due to the high concentration of PE outside the MLV). Microscopically, the PE molecules may deform the bilayer, weaken the cohesion among the organized DDAB molecules, and create defects; hence membrane subunits may temporally be separated, allowing the passage of PE molecules (as observed with lipid vesicles in presence of surfactant) [62-64]. We finally note that the penetration of a polymer across a lipid bilayer has been recently observed experimentally,[65-66] in concordance with our experimental findings.

## 4 Conclusions

In summary, we have provided experimental data on the kinetics of formation of synthetic charged lipids/polyelectrolyte complexes. By using multilamellar vesicles and a high polyelectrolyte concentration gradient, we were able to visualize by confocal imaging the progressive binding of the charged bilayers as they interact with oppositely charged polyelectrolyte. Although PE/lipid interactions have previously been visualized on the nanometer scale by atomic force microscopy [67-69], our experiments constitute, to the best of our knowledge, one of the first observations on the micrometer scale. We have described the microscopic mechanisms at play and have provided quantitative measurements, which support our physical picture. The key parameter for this novel observation is to expose MLV to a strong gradient of PE. We indeed have demonstrated that a weak gradient induced radically different morphological transitions. Our description of a gradual binding process of charged bilayers induced by oppositely charged polyelectrolyte may shed some light for understanding the more complicated cell membrane behaviors induced by different kinds of charged proteins. Finally, we have also shown that a strong gradient induced eventually the spontaneous evolution of a MLV towards a hollow capsule. Our simple approach may be useful in designing a class of soft composite polyelectrolyte/lipid shell for applications for drug delivery or controlled drug release.

**Acknowledgment**. We acknowledge financial support from the CNRS-CEA-DFG-MPIKG Network "Complex fluids: from 3 to 2 dimensions" and from the European Network of Excellence "SoftComp" (NMP3-CT-2004-502235). We thank G. Porte for fruitful discussions.

**Supporting Information Available:** Materials and Methods; Movie showing the initial process of budding formation as a MLV interacts with a concentrated PE solution (30%W/W).

TOC graphic

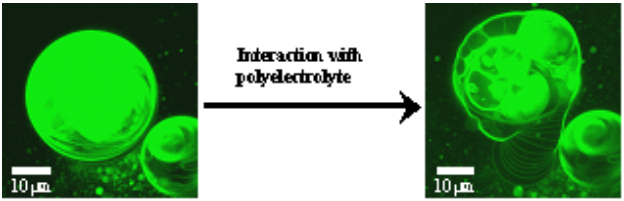